%% file: main.tex
\documentclass[francais]{gretsi}

\usepackage[utf8]{inputenc}

\usepackage[T1]{fontenc}

\input{loadpackage}

\begin{document}

    \titre{Apprentissage auto-supervisé pour la reconstruction de phase}

    \auteurs{
      \auteur{Victor}{Sechaud}{victor.sechaud@ens-lyon.fr}{1}
      \auteur{Patrice}{Abry}{patrice.abry@ens-lyon.fr}{1}
      \auteur{Laurent}{Jacques}{laurent.jacques@uclouvain.be}{2}
      \auteur{Julián}{Tachella}{julian.tachella@ens-lyon.fr}{1}
    }
    \affils{
      \affil{1}{CNRS, ENS de Lyon, LPENSL, UMR5672, 69342, Lyon Cedex 07, France}
      \affil{2}{UCLouvain, ICTEAM, Louvain-la-Neuve, Belgium}
    }

    \resume{Ces dernières années, les réseaux neuronaux profonds se sont imposés comme solution pour les problèmes inverses en imagerie.
    Ces réseaux sont généralement entraînés à l'aide de paires d'images : l'une dégradée et l'autre en haute qualité, ces dernières étant appelées « vérités terrain ».
    Toutefois, en imagerie médicale, le manque de données de références limite l'apprentissage supervisé.
    Des avancées récentes ont permis de reconstruire des images à partir des seules données de mesure, supprimant ainsi le besoin de vérités terrain.
    Cependant, ces méthodes restent limitées aux problèmes linéaires, excluant les processus non linéaires comme la reconstruction de phase.
    Nous proposons une méthode auto-supervisée qui surmonte cette difficulté dans le cas de reconstruction de phase en utilisant l'invariance naturelle des images aux translations.}

    \abstract{In recent years, deep neural networks have emerged as a solution for inverse imaging problems.
    These networks are generally trained using pairs of images: one degraded and the other of high quality, the latter being called ‘ground truth’.
    However, in medical and scientific imaging, the lack of fully sampled data limits supervised learning.
    Recent advances have made it possible to reconstruct images from measurement data alone, eliminating the need for references.
    However, these methods remain limited to linear problems, excluding non-linear problems such as phase retrieval.
    We propose a self-supervised method that overcomes this limitation in the case of phase retrieval by using the natural invariance of images to translations.
    }

    \maketitle


    \section{Introduction}
    Les problèmes inverses jouent un rôle essentiel dans diverses applications en ingénierie et en sciences, telles que la tomographie et l'IRM~\cite{lustig_sparse_2007}.
    Ces problèmes impliquent la reconstruction d'un signal inconnu $\x \in \X \subseteq \R^n$ à partir de mesures observées $\y \in \Y \subseteq \R^m$, modélisées par l'équation :
    \[
    \y = h(\x) + \boldsymbol{\epsilon},
    \]
    où $h : \R^n \mapsto \R^m$ est l'opérateur direct, et $\boldsymbol{\epsilon} \in \R^m$ représente le bruit.
    Ces problèmes sont souvent mal posés en raison de deux facteurs principaux : l'opérateur direct $h$ peut entraîner une perte d'information (par exemple, lorsque $m < n$) et le bruit introduit une incertitude supplémentaire.

    Pour y remédier, des informations préalables sur le signal sont généralement incorporées, telles que des hypothèses de constances par morceaux~\cite{rudin_nonlinear_1992}.
    Si les a priori peuvent permettre des reconstructions robustes, en choisir un approprié est difficile, car un a priori inexact peut biaiser la solution et donc ne pas représenter le signal sous-jacent de manière efficace.

    Une autre solution consiste à apprendre directement la fonction de reconstruction qui approxime l'opérateur inverse de $h$.
    Les méthodes d'apprentissage supervisé y parviennent en entrainant un réseau de neurones $f_{\boldtheta} : \Y \to \X$.
    Les paramètres du réseau $\boldtheta \in \R^p$ sont optimisés en minimisant l'erreur quadratique moyenne (EQM) sur un ensemble de données de signaux de vérité de terrain appariés et leurs mesures, $\left\{(\x_i, \y_i)\right\}_{i \in I}$ :
    \begin{equation}
        \boldsymbol{\hat{\theta}} = \argmin\limits_{\boldtheta} \sum\limits_{i \in I} \|f_{\boldtheta}(\y_i) - \x_i\|^2. \label{mse}
    \end{equation}
    Quoique cette approche donne fréquemment de bons résultats, elle présente deux inconvénients majeurs :
    \textit{(i)}
    disponibilité des données d'entrainement : l'obtention d'un ensemble de données d'entraînement complet avec des vérités terrain précis peut être difficile ou peu pratique, en particulier dans des domaines tels que l'imagerie scientifique~\cite{belthangady_applications_2019}.
    \textit{(ii)}
    Décalage de la distribution : même avec un ensemble de données approprié, les performances peuvent se dégrader en cas de décalage entre la distribution des données d'apprentissage et celle des données réelles.

    L'apprentissage auto-supervisé offre une alternative qui surmonte ces défis~\cite{belthangady_applications_2019}.
    Contrairement aux approches supervisées, il ne nécessite pas d'ensembles de données contenant des signaux de référence.
    Au lieu de cela, les méthodes auto-supervisées peuvent être entrainées directement sur les données de mesure, ce qui est particulièrement avantageux lorsque les données de vérité terrain sont coûteuses ou indisponibles.

    En outre, en évitant de s'appuyer sur les vérités terrain, l'apprentissage auto-supervisé réduit le risque de décalage de la distribution.

    Cependant, l'apprentissage uniquement à partir de mesures présente des difficultés particulières, notamment lorsque l'opérateur direct n'est pas inversible.
    Pour les opérateurs linéaires, l'apprentissage auto-supervisé devient possible en supposant que l'espace de signal sous-jacent présente une invariance par rapport à des transformations spécifiques, telles que des rotations ou des translations~\cite{chen_equivariant_2021}.
    Cette approche peut également être étendue à certains problèmes non linéaires, tels que la reconstruction de signaux fortement quantifiés, par exemple, des mesures à $1$-bit~\cite{tachella_binary_2023} ou le declipping~\cite{sechaud_equivariance_2024}.

    Dans ce travail, nous proposons un cadre d'apprentissage auto-supervisé qui applique l'équivariance pour les translations.
    Ce cadre est appliqué aux problèmes de reconstruction de phase, démontrant par des expériences qu'il permet d'obtenir des performances comparables aux méthodes entièrement supervisées.

    Nos principales contributions sont les suivantes :
    \begin{enumerate}
        \item Fonction de perte auto-supervisée : Nous introduisons une fonction de perte pour l'entraînement des réseaux neuronaux dans les tâches de reconstruction de phase qui ne nécessite pas la connaissance de vérité terrain.
        \item Nous montrons que la méthode auto-supervisée proposée peut être aussi performante que l'approche supervisée.
    \end{enumerate}

    \vspace{-5mm}
    \section{État de l'art}
    \vspace{-2mm}
    \textbf{Apprentissage auto-supervisé -}
    Dans le contexte de la restauration de données (comme le débruitage ou le défloutage), les méthodes auto-supervisées permettent de former des réseaux de reconstruction en utilisant uniquement des données de mesure~\cite{chen_equivariant_2021}.
    Les méthodes existantes peuvent être classées en deux catégories : celles qui traitent le bruit du problème, par exemple les méthodes Noise2X~\cite{lehtinen_noise2noise_2018, krull_noise2void-learning_2019}, et celles qui traitent les opérateurs non inversibles~\cite{tachella_sensing_2023}.
    Le dernier ensemble de méthodes se concentre principalement sur le cas de l'opérateur de mesure linéaire où la perte d'information est associée à son espace nul.
    Il est possible de remédier à ce manque d'information en
    \textit{(i)} en accédant aux mesures de plusieurs opérateurs ayant des espaces nuls différents~\cite{tachella_sensing_2023}, ou
    \textit{(ii)} en supposant que la distribution du signal soit invariante par rapport à un ensemble de transformations telles que les translations ou les rotations, une méthode connue sous le nom d'imagerie équivariante (EI)~\cite{chen_equivariant_2021}.
    Le cadre de l'EI est bien étudié pour les problèmes linéaires inverses, à la fois en termes d'applications~\cite{scanvic_self-supervised_2024, chen_equivariant_2021} et de théorie~\cite{tachella_sensing_2023}.
    À notre connaissance, les seuls cas de problèmes inverses non linéaires dans le cadre de l'EI sont l'acquisition compressive à un bit~\cite{tachella_binary_2023} et la désaturation~\cite{sechaud_equivariance_2024}.

    \textbf{Reconstruction de phase -} \label{sec: phase}
    Les méthodes classiques de reconstruction de phase, telles que l'algorithme de Gerchberg-Saxton~\cite{gerchberg_practical_1972}, reposent sur un raffinement itératif, mais sont limitées par une convergence lente et une sensibilité au bruit.
    Les approches basées sur des modèles répondent à ces limitations en incorporant des techniques de régularisation telles que la sparsité~\cite{ohlsson_conditions_2014} pour améliorer la qualité de la reconstruction ; cependant, leurs performances dépendent fortement de la précision des a priori supposés.
    Les techniques d'optimisation convexe, notamment PhaseLift~\cite{candes_phase_2015} et PhaseMax~\cite{goldstein_phasemax_2018},
    reformulent la reconstruction de phase comme un problème convexe en se plaçant dans l'espace de matrice ou en utilisant la programmation semi-définie.
    Bien que ces méthodes offrent des garanties de récupération sous certaines conditions, elles nécessitent des calculs intensifs et des ratios d'échantillonnage élevés.
    Enfin, les méthodes d'apprentissage, principalement basées sur des réseaux de neurones, ont récemment émergé comme une alternative efficace~\cite{dong_phase_2023}. 

    \vspace{-4mm}
    \section{Reconstruction de phase}
    \vspace{-1mm}
    \subsection{Formulation du problème}
    Soit \(\A \in \C^{m \times n}\) une matrice dont les lignes sont notées \(\{\boldsymbol{a}_j\}_{1 \leq j \leq m}\).
    Nous définissons la formulation directe \(\y = h(\x)\) comme suit :
    \[
        y_j = |\boldsymbol{a}_j^\top \x|^2, \quad \forall j \in \{1, \dots, m\},
    \]
    ou sous forme matricielle :
    \[\y = |\A\x|^2,\]
    pour tout \(\x \in \X\).
    Le modèle \(\X \in \C^n\) correspond à l'ensemble des signaux que nous cherchons à reconstruire, généralement bien plus petit que \(\C^n\).
    L'opérateur \(\A\) représente la composante linéaire de la physique d'acquisition et est souvent une matrice aléatoire ou une matrice de transformation de Fourier~\cite{shechtman_phase_2015}.
    Le rapport \( \alpha = m/n \) représente le taux d’échantillonnage.
    Un $\alpha$ petit signifie une compression élevée des mesures, rendant la reconstruction plus difficile, mais l'acquisition moins couteuse.

    \vspace{-2mm}
    \subsection{Hypothèse d'invariance}\label{sec: invariance}
    Dans notre travail, nous considérons un modèle $\X$ invariant à certaines transformations, comme les translations ou les rotations.
    Mathématiquement, pour un groupe de transformations \(G\), cette invariance s'écrit :
    \begin{equation}
        \tg\X = \X \quad \text{pour tout } g \in G,
    \end{equation}
    où \(\tg\) est l'opérateur de transformation linéaire associé à \(g \in G\).
    Cette hypothèse est naturelle pour différentes transformations et nombreux signaux : l'ensemble des images naturelles est invariant aux rotations et translations.
    La raison de considérer cette hypothèse est que l'invariance aux transformations \(\tg\) nous donne accès aux ensembles \(\Y_g = h(\tg\X)\) pour tout \(g \in G\).
    En effet, nous pouvons voir \(\Y\) comme un ensemble de mesures associées au modèle \(\X\) pour différents opérateurs directs \(h_g(\cdot) := h(\tg\cdot)\) grâce au fait que
    \[\Y_g = | \A \tg\X|^2 = | \A\X|^2 \quad \textrm{si } \quad \tg\X = \X.\]
    Cela peut fournir des informations supplémentaires au-delà du noyau de \(\A\), nous permettant d'apprendre l'ensemble des signaux \(\X\) à partir des ensembles de mesures \(\Y_g\).
    Toutefois, pour que cette approche soit efficace, la transformation \(\tg\) doit modifier le noyau de \(\A\), c'est-à-dire que $\ker(\A) \neq \ker( \A \tg)$ pour tout $g \in G$.
    Une condition nécessaire est que l'opérateur $\A$ ne commute pas avec les transformations du groupe $G$~\cite{tachella_sensing_2023}.

    \vspace{-2mm}
    \subsection{Metrique}
    La reconstruction du signal \( \x \in \C^n \) à partir des observations \( \y = |\A\x|^2 \) n'est pas unique.
    En effet, toute transformation de la forme \( \x \mapsto e^{i\varphi} \x \), où \( \varphi \) est un réel, préserve les mesures d’intensité :
    \[
    |\A(e^{i\varphi}\x)|^2 = |\A\x|^2.
    \]
    Par conséquent, toute méthode de reconstruction ne peut estimer \( \x \) qu’à une phase globale près.
    Pour évaluer la qualité de reconstruction, nous utilisons donc la similarité cosinus, notée \( \textrm{CS} ( \cdot, \cdot ) \), qui est insensible à ces variations de phase.
    Cette métrique est définie par :
    \begin{equation}
        \textrm{CS}(\x, \hat{\x}) = \frac{|\x^* \hat{\x}|}{\|\x\|_2 \|\hat{\x}\|_2},
        \label{eq:cosinus}
    \end{equation}
    où \( \x^* \) désigne l'adjoint de \( \x \).
    Une similarité cosinus de \( 1 \) indique  indique que la reconstruction est parfaite, c'est-à-dire qu'il existe une phase (et éventuellement un facteur d’échelle) tel que
    \[
        \hat{\mathbf{x}} = r\, e^{i\varphi}\mathbf{x}, \quad \text{avec } r \in \mathbb{R} \text{ et } \varphi \in \mathbb{R}.
    \]
    Au contraire, une valeur proche de \( 0 \) indique une reconstruction aléatoire.
    Le facteur d’échelle \( r \) n'est pas problématique car il peut facilement être connue en remarquant que
    \[
        |\A \boldsymbol{\hat{x}}|^2 = |\A re^{i \varphi} \x|^2 = r^2|\A\x|^2 = r^2\y.
    \]
    Dans la suite, pour parler d'égalité à une phase globale près, nous utiliserons le symbole \( \eqphase \).
    Ainsi, \( \hat{\x} \eqphase \x \) signifie
    \[
        \exists \varphi \in \R, \textrm{ tel que }  \hat{\x} = e^{i \varphi} \x.
    \]

    \vspace{-5mm}
    \section{Apprentissage auto-supervisé}

    \textbf{Fonction de perte de consistence des mesures (CM) -}
    Une méthode de reconstruction efficace doit respecter la consistence des mesures, c'est-à-dire que pour toute mesure $\y$, nous avons
    \[
    h(f_{\boldtheta}(\y)) = \y.
    \]
    Garantir la consistence de mesure est essentiel, car cela assure que les estimations restent fidèles aux observations.
    Cependant, cela n'implique pas nécessairement l'unicité de la solution, car plusieurs solutions peuvent satisfaire les mêmes mesures.
    Une fonction de perte naturelle pour imposer cette consistence est :
    \begin{equation}
        \mathcal{L}_\textrm{CM}(\boldtheta) = \sum\limits_{i\in I} \|\y_i - h\left(f_{\boldtheta}(\y_i)\right)\|^2. \label{eq: loss intensity}
    \end{equation}
    Cette fonction de perte est aussi connue dans le domaine de reconstruction de phase comme la perte d'intensité.
    Nous pouvons également considérer une variante de cette perte, la perte d'amplitude :
    \begin{equation}
        \mathcal{L}_\textrm{A}(\boldtheta) = \sum\limits_{i\in I} \left\|\sqrt{\y_i} - \sqrt{h\left(f_{\boldtheta}(\y_i)\right)}\right\|^2. \label{eq: loss amplitude}
    \end{equation}
    Bien que ces deux fonctions aient le même minimum global pour un \(f_{\theta}\) suffisamment flexible, des observations empiriques suggèrent qu'il est parfois préférable d'utiliser la perte d'amplitude, particulièrement dans le cas de signaux bruités~\cite{yeh_experimental_2015}.

    \noindent \textbf{Fonction de perte d'équivariance - }
    Pour apprendre efficacement la phase, nous utilisons l'hypothèse d'invariance du modèle définie dans la section~\ref{sec: invariance}, et allons ainsi au-delà des limitations imposées par la consistence des mesures seule.
    Pour exploiter sur le réseau de reconstruction cette propriété d'invariance, nous remarquons que, puisque \(\tg\x \in \X\), la fonction \(f\) doit être capable de reconstruire à la fois \(\x\) et \(\tg\x\) pour tout \(\x \in \X\) et \(g \in G\), c'est-à-dire :
    \begin{equation}
    f\left(h(\tg\x)\right) \eqphase \tg\x \quad \text{et} \quad f(h(\x)) \eqphase \x.
    \end{equation}
    Ainsi, nous voulons avoir :
    \begin{equation}
    f(h(\tg\x)) \eqphase \tg f(h(\x)) \textrm{ pour tout } \x \in \X, \textrm{ et } g \in G.
    \end{equation}
    La composition de \(f\) avec \(h\) doit alors être équivariante par rapport au groupe \(G\).
    Pour garantir cela, nous proposons une fonction de perte qui impose cette propriété d'équivariance :
    \begin{equation}
        \mathcal{L}_{\textrm{EI}}(\boldtheta) =  - \sum_{i\in I}\sum_{g\in G}\textrm{CS}\Big(\tg f_{\boldtheta}\left(\y_i\right), f_{\boldtheta}\big(h(\tg f_{\boldtheta}\left(\y_i\right))\big)\Big). \label{loss ei}
    \end{equation}
    L'utilisation de la CS dans la fonction de perte ne garantit pas des facteurs d'échelle $r$ corrects entre $\tg f_{\boldtheta}\left(\y_i\right) $ et $ f_{\boldtheta}\big(h(\tg f_{\boldtheta}\left(\y_i\right))\big)$ ; c'est pourquoi il est nécessaire de l'associer à une fonction de perte de consistence des mesures.

    La fonction de perte finale est donc une combinaison des pertes de consistence des mesures et d'équivariance :
    \begin{equation}
        \mathcal{L}(\boldtheta) = \mathcal{L}_\textrm{A}(\boldtheta) + \lambda \mathcal{L}_{\textrm{E}}(\boldtheta). \label{eq: loss}
    \end{equation}
    Le paramètre \(\lambda\) est un hyperparamètre positif qui contrôle le compromis entre les deux pertes.
    La fonction de reconstruction peut être basée sur n'importe quelle architecture de réseau de neurones.
    Pour nos expériences, nous avons choisi d'utiliser un U-Net~\cite{ronneberger_u_2015}.
    
    \vspace{-4mm}
    \section{Expériences}
\begin{figure}
        \includegraphics{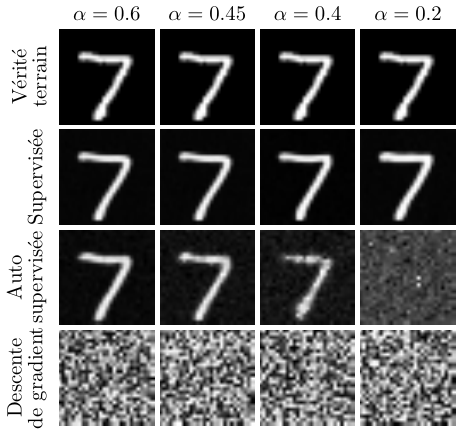}
        \vspace*{-5mm}
        \caption{Images reconstruites pour les méthodes supervisée, auto-supervisée et pour une résolution par descente de gradient.}
        \label{fig: mnist}
    \end{figure}

    Dans cette section, nous décrivons la méthodologie adoptée pour tester et comparer notre approche.
    L'ensemble des expériences ont été réalisées à l'aide de la bibliothèque open-source DeepInv~\cite{tachella_deepinverse_2023} sur le cluster de machine du Centre Blaise Pascal à l'ENS de Lyon~\cite{quemener_sidus_2013}.
    Tout d'abord, nous construisons un jeu de données en utilisant le dataset MNIST.
    À partir des images initiales \( \x_0 \in \R^n \), où $n = 28 \times 28$, nous générons unes images complexes \( \x = e^{i \x_0} \in \C^n\) puis nous calculons les mesures \( \y = |\A\x|^2 \).
    La matrice complexe \( \A \) est une matrice aléatoire de dimensions \( m \times n \), dont chaque élément \( a_{k,l} \) suit une distribution normale centrée, donnée par :
    \[
        a_{k,l} \sim \mathcal{N}(0, 1/2m) + i\mathcal{N}(0, 1/2m).
    \]
    Ces matrices servent à modéliser des milieux diffusants, utile en imagerie optique~\cite{liutkus_imaging_2014}.
    \begin{figure}[t]
        \includegraphics{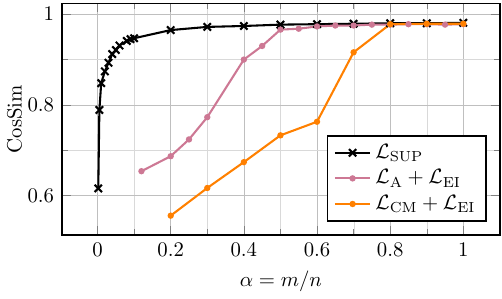}
        \vspace*{-6mm}
        \caption{Cosinus similarité en fonction du taux d'échantillonnage $\alpha$ pour la méthode supervisée et les méthodes auto-supervisées.}
        \label{fig: slice_phase}
    \end{figure}
    Nous évaluons notre méthode pour plusieurs valeurs du ratio d'échantillonnage \( \alpha \) pour le groupe de transformations des translations, c'est à dire $\tg$ agit sur $\x$ comme un shift des pixels.
    Les méthodes classiques de reconstruction de phase, vus en section~\ref{sec: phase} comme PhaseLift et PhaseMax nécessitent un taux d'échantillonnage élevé, $\alpha > 2$.
    Cependant elles ne sont pas restreintes sur un modèle $\X$ et peuvent reconstruire sur l'espace $\C^n$ tout entier.
    Une comparaison avec ces méthodes ne serait donc pas pertinente ; c'est pourquoi nous nous limitons à une comparaison avec la méthode supervisée.
    Nous modifions la fonction de perte~\ref{mse} pour l'adaptée à la reconstruction de phase :
    \[
        \mathcal{L}_\textrm{SUP}(\boldtheta) = - \sum\limits_{i\in I} \textrm{CS}\big(\x_i, f_{\boldtheta}(\y_i)\big),
    \]
    Pour $k \in \left\{1, \dots, m \right\}$ et $l \in \left\{1, \dots, n \right\}$.
    Les résultats sont représenté dans la figure~\ref{fig: slice_phase}.
    Enfin, nous corrigeons la phase globale de \( f(\y) \) pour l'aligner avec \( \x \) afin d’assurer une représentation visuelle cohérente.
    Un exemple de reconstruction pour différente valeur de $\alpha$ est donné dans la figure~\ref{fig: mnist}.
    On remarque que pour des valeurs de $\alpha$ supérieures à $0.5$, la méthode auto-supervisée donne des résultats comparables à la méthode supervisée.
    En revanche, cette dernière performe nettement mieux pour des valeurs de $\alpha$ faible ($\alpha < 0.3$).
    On observe également que méthode entrainé avec la perte d'intensité~\ref{eq: loss intensity} en tant que CM permet une compression moindre que celle entrainé avec la perte d'amplitude~\ref{eq: loss amplitude}. \\
    \textbf{Paramètre d'apprentissage - } Nous avons utilisé un réseau U-Net comportant 4 niveaux de descente et 4 niveaux de montée.
    Le paramètre $\lambda$ de la fonction de perte~\ref{eq: loss} est fixé à $1$ dans toutes les expériences.
    Pour des raisons computationnelles lors du calcul de la fonction de perte EI~\ref{loss ei}, nous sommons sur deux translations par image, plutôt que sur l'ensemble des transformations du groupe $G$.
    L'optimisation a été effectuée à l'aide de l'algorithme Adam avec un taux d'apprentissage de $5e^{-5}$.
    Le modèle a été entraîné sur $15$ epochs avec une taille de batch de $5$, sur un tier des images de MNIST.

    \vspace{-3mm}
    \section{Conclusion}
    \vspace{-1mm}
    Nous avons présenté une approche auto-supervisée qui tire parti de l'invariance pour résoudre le problème de la reconstruction de phase.
    Nos expériences préliminaires montrent que notre méthode atteint des performances comparables à celles de l'approche supervisée dans le cadre de l'invariance par translation.
    Ce travail met en évidence de nouvelles possibilités de solutions basées sur l'apprentissage pour les problèmes inverses non linéaires.
    D'autres travaux peuvent être réalisés pour étendre cette approche à divers groupes de transformations ou d'autre type de signaux, tout en proposant un cadre théorique assurant la faisabilité de la reconstruction.
    
    \vspace{-3mm}
    \section{Remerciements}
    \vspace{-1mm}
    Ce travail a été financé par le projet ANR-20-CE40-0001-01.
    La recherche de Laurent Jacques est en partie financée par le FRS-FNRS (QuadSense, T.0160.24).
    
    \vspace*{-3mm}
    {\footnotesize
    \bibliography{biblio}}

\end{document}

%% file: loadpackage.tex
\usepackage{amsmath,amsfonts,amsthm}
\usepackage{bm}
\usepackage{bbm}
\usepackage{stmaryrd}
\usepackage{array}
\usepackage{verbatim}

\usepackage{url}
\usepackage{cite}
\usepackage{cleveref} 

\usepackage{graphicx} 
\usepackage[caption=false,font=normalsize,labelfont=sf,textfont=sf]{subfig}

\usepackage{tikz}
\usepackage{tikz-3dplot}
\usetikzlibrary {patterns.meta}
\usetikzlibrary{patterns}
\usepackage{standalone}
\usepackage{pgfplots}
\pgfplotsset{compat=1.18} 
\usepackage{fp}

\usepackage{xcolor}
\definecolor{rose_ens}{RGB}{205,120,148}
\definecolor{bleu_ens}{RGB}{30,60,80}
\definecolor{darkgray176}{RGB}{176,176,176}
\definecolor{gray}{RGB}{128,128,128}
\definecolor{slategray100100150}{RGB}{100,100,150}

\newcommand*{\R}{\mathbb{R}}
\newcommand*{\C}{\mathbb{C}}
\newcommand*{\X}{\mathcal{X}}
\newcommand*{\Y}{\mathcal{Y}}
\newcommand*{\x}{\boldsymbol{x}}
\newcommand*{\y}{\boldsymbol{y}}

\newcommand*{\A}{\boldsymbol{A}}
\newcommand*{\boldtheta}{\boldsymbol{\theta} }
\newcommand*{\tg}{\boldsymbol{T}_g}
\newcommand*{\eqphase}{\overset{\varphi}{=}}

\DeclareMathOperator*{\argmin}{\arg\,min}

\usepackage{lipsum} 
